\documentclass{article}

\usepackage{amsmath} 
\usepackage{physics} 

\usepackage{xcolor}
\usepackage{graphicx}
\usepackage{subfigure}

\usepackage{booktabs} 
\usepackage{array} 
\usepackage{multirow} 
\usepackage{multicol} 
\usepackage{adjustbox} 
\usepackage{makecell} 
\usepackage{longtable} 
\usepackage{caption} 
\usepackage{algorithm}
\usepackage{algorithmicx}
\usepackage{algpseudocode}

\usepackage{cite}
\makeatletter
\def\@cite#1#2{\textsuperscript{[{#1\if@tempswa , #2\fi}]}} 
\makeatother

\usepackage{amssymb} 
\makeatletter
\newcommand{\Rmnum}[1]{\expandafter\@slowromancap\romannumeral #1@} 
\makeatother

\definecolor{goodorange}{RGB}{225,125,0}
\definecolor{goodgreen}{RGB}{5,130,5}
\definecolor{goodred}{RGB}{220,50,25}
\definecolor{goodblue}{RGB}{30,144,255}
\definecolor{jpcablue}{RGB}{0,84,166}
\definecolor{OliveGreen}{RGB}{5,100,5}

\usepackage{ifthen}
\usepackage{xspace}

\newcommand{\note}[2]{
\ifthenelse{\equal{#1}{Z}}{
\colorbox{jpcablue}{\textcolor{white}{\footnotesize \fontfamily{phv}\selectfont #1}}
    \textcolor{jpcablue}{{\footnotesize \fontfamily{phv}\selectfont #2}}\xspace
}{}
\ifthenelse{\equal{#1}{D}}{
\colorbox{goodorange}{\textcolor{white}{\footnotesize \fontfamily{phv}\selectfont #1}}
    \textcolor{goodorange}{{\footnotesize \fontfamily{phv}\selectfont #2}}\xspace
}{}
\ifthenelse{\equal{#1}{Y}}{
\colorbox{goodred}{\textcolor{white}{\footnotesize \fontfamily{phv}\selectfont #1}}
    \textcolor{goodred}{{\footnotesize \fontfamily{phv}\selectfont #2}}\xspace
}{}
}

\begin{document}
\title{Stochastic resolution of identity to CC2 for large systems: Excited-state gradients and derivative couplings}
\author{Chongxiao Zhao\textsuperscript{1,2,a)}, Chenyang Li\textsuperscript{3,b),*}, and Wenjie Dou\textsuperscript{1,2,b),*}}
\date{}
\maketitle 

\begin{center}
    1.Department of Chemistry, School of Science, Westlake University, Hangzhou, Zhejiang 310024, China \\
    2.Institute of Natural Sciences, Westlake Institute for Advanced Study, Hangzhou, Zhejiang 310024, China \\
    3.Key Laboratory of Theoretical and Computational Photochemistry, Ministry of Education, College of Chemistry, Beijing Normal University, Beijing 100875, China \\
    \textsuperscript{a)}Email: zhaochongxiao@westlake.edu.cn \\
    \textsuperscript{b)}Authors to whom correspondence should be addressed: chenyang.li@bnu.edu.cn and douwenjie@westlake.edu.cn \\
\end{center}

\begin{abstract}
Excited-state gradients and derivative couplings are critical for simulating excited-state dynamics.
However, their calculations are very expensive within the coupled-cluster framework due to the steep scaling.
In this work, we present two implementations of stochastic resolution of identity to CC2 (sRI-CC2) for excited-state analytical gradients and derivative couplings.
The first method employs sRI for both Coulomb and exchange terms, reducing the formal scaling to cubic.
However, this method has a significant stochastic noise. 
Consequently, we introduce a substitute, termed partial sRI-CC2, which applies sRI selectively to the exchange terms only.
The partial sRI-CC2 shows a quartic scaling with a modest prefactor, rendering it a practical alternative.
Compared to conventional RI-CC2, the partial sRI-CC2 can handle systems with hundreds or even thousands of electrons.
This work is an extension to our previous implementation of sRI-CC2 method and provides essential ingredients for large-scale nonadiabatic dynamics.
\end{abstract}

\section*{\Rmnum{1}. INTRODUCTION}

\hspace{1em} 
Over the past decades, the CC2 model has been a well-established method for the accurate calculation of electronic structure properties.
The integration of various techniques, such as pair natural orbital (PNO) approach,\cite{ahlrichs1968direct,neese2009efficient,helmich2011local,helmich2013pair,schmitz2014explicitly} local correlation approximation,\cite{pulay1983localizability,hampel1996local,schutz2001low,kats2006local}
tensor hypercontraction (THC),\cite{gray2021hyper,hohenstein2012tensor,hohenstein2013quartic,hohenstein2013tensor,sacchetta2025efficient,lee2021even}
spin-component scaling (SCS)\cite{grimme2003improved,hellweg2008benchmarking,tajti2019accuracy} and scaled opposite-spin (SOS) modifications,\cite{jung2004scaled,winter2011scaled,sacchetta2025efficient} and others, significantly reduces the original fifth-order computational cost of CC2
and substantially broadens the applicability of CC2 to larger molecular systems.
However, most improvements yield only an order of magnitude reduction and a smaller prefactor.
Consequently, the application of CC2 to very large systems remains a significant computational challenge.

Among the potential approximation methods, a stochastic variant of resolution of identity (sRI) presents a compelling option.
Based on the prevalent RI approach,\cite{feyereisen1993use,eichkorn1995auxiliary,bernholdt1996large} this technique employs randomly generated stochastic orbitals to further decouple the four-index electron repulsion integrals (ERIs).
Combined with Laplace transform,\cite{almlof1991elimination,haser1992laplace,haser1993moller} the sRI approximation is very efficient in evaluating MP2-energy-like intermediates, which is ubiquitous in common electronic structure methods.
Therefore, the sRI technique gains rapid popularity in many implementations, such as DFT,\cite{baer2013self,neuhauser2014communication,gao2015sublinear,neuhauser2016stochastic,bradbury2023deterministic}, MP2, \cite{neuhauser2013expeditious,ge2014guided,takeshita2017stochastic} GF2\cite{neuhauser2017stochastic,takeshita2019stochastic,dou2019stochastic,dou2020range,mejia2024convergence}
and etc.\cite{rabani2015time,lee2020stochastic}
Recently, we introduce sRI to CC2 ground and excited-state energy,\cite{zhao2024stochastic_gs,zhao2024stochastic_ex} oscillator strength and ground-state analytical gradient,\cite{zhao2025stochastic} achieving a steep scaling reduction from $O(N^5)$ to $O(N^3)$ or $O(N^4)$.
This reduced computational cost enables CC2 calculations on systems with hundreds or even thousands of electrons.
To complete the sRI-CC2 series and facilitate its application in dynamics simulations, the next step falls on excited-state analytical gradients and derivative couplings.

The analytical energy derivative of excited states is essential for determining equilibrium geometries and conducting molecular dynamics simulations.
The explicit formulations of CC2 excited-state gradient were first reported by H{\"a}ttig $et$ $al$.\cite{kohn2003analytic,winter2012quartic}
Afterwards, several improvements were carried out.\cite{Ledermller2013local,Ledermller2014local,Heuser2018analytical}
The derivative coupling is also critical for dynamics simulations, and the CC2 method represents a promising approach for their calculation.
However, the situation of CC2 derivative couplings is more complex, and their feasibility was a subject of debate two decades ago.\cite{hattig2005structure,kohn2007can}
Due to the lack of Hermitian symmetry, conical intersections between states spanning the same symmetry would encounter complex energy and result in a defective intersection.
This challenge was recently addressed by Koch $et$ $al$. through the development of a modified CC2 theory, termed similarity constrained CC2 (SCC2).\cite{kjonstad2023communication,kjonstad2024coupled,stoll2025similarity} 
Their approach is demonstrated to correctly describe conical intersections within the coupled cluster framework.\cite{kjonstad2017resolving,kjonstad2019orbital,rossi2025generalized}

CC2 has been implemented in various software packages including Psi4,\cite{smith2020psi4} COLUMBUS,\cite{lischka2020generality} Pyscf,\cite{sun2020recent}.
This work is mainly developed using the Q-Chem package\cite{epifanovsky2021software,paran2022spin,paran2024performance,alessio2024coupled} and presents two implementations of sRI-CC2 for the computation of excited-state analytical gradients and derivative couplings.
The first, designated the complete sRI-CC2 method, is implemented with cubic scaling.
However, its practical application is severely constrained by the requirement for a large number of stochastic orbitals.
To overcome this limitation, we introduce a quartic-scaling variant, termed partial sRI-CC2, which employs sRI selectively within the exchange terms.
This partial implementation demonstrates superior accuracy and its reduced scaling enables the calculations of systems with hundreds or even thousands of electrons.

This paper is organized as follows:
In Sec. II, we present the theoretical framework and detailed algorithms for excited-state analytical gradient and derivative coupling at the sRI-CC2 level.
This section also examines the differences between RI and sRI approximations and outlines the justification for our partial sRI-CC2.
Sec. III evaluates the numerical accuracy of the sRI-CC2 method and compares its computational cost to the RI-CC2 method.
Sec. IV gives a conclusion.

\section*{\Rmnum{2}. THEORY}

\hspace{1em} 
The notation defined in Table \ref{tab:notation} is used throughout this work.
All values in the final column scale proportionally with the system size $N$.
\begin{table}[htbp]
    \caption{Summary of notations in the following equations.} 
    \begin{adjustbox}{center} 
        \begin{tabular}{lcc} 
            \toprule[0.3pt]
            \specialrule{0em}{0.3pt}{0.3pt} 
            \midrule
            \makebox[0.3\textwidth][l]{Item} & \makebox[0.3\textwidth][c]{Function or indices} & \makebox[0.15\textwidth][c]{Total number} \\ 
            \midrule
            AO Gaussian basis functions & $\chi _\alpha (r_1)$, $\chi _\beta (r_1)$, $\chi _\gamma (r_1)$, $\chi _\delta (r_1)$, $\cdots$ & $N_{AO}$ \\
            Auxiliary basis functions & $P$, $Q$, $R$, $S$, $\cdots$ & $N_{aux}$ \\
            General sets of AOs & $\alpha $, $\beta $, $\gamma $, $\delta $, $\cdots$ & $N_{ao}$ \\
            General sets of MOs & $p$, $q$, $r$, $s$, $\cdots$ & $N_{mo}$ \\
            Occupied (active) MOs & $i$, $j$, $k$, $l$, $\cdots$ & $N_{occ}$ \\
            Unoccupied (virtual) MOs & $a$, $b$, $c$, $d$, $\cdots$ & $N_{vir}$ \\
            \midrule
            \specialrule{0em}{0.3pt}{0.3pt}
            \bottomrule[0.3pt]
        \end{tabular}
    \end{adjustbox}
    \label{tab:notation}
\end{table}

\subsection*{A. CC2 theory}

\hspace{1em} 
In the CC2 formulation,\cite{christiansen1995second} the $T_1$-transformed Hamiltonian $\hat H$ provides a convenient framework
\begin{equation}
    \hat H = e^{-T_1} H e^{T_1}
\end{equation}
The cluster operators $T_n = \sum_{\mu_n} {t_{\mu_n} \tau_{\mu_n}}$ consist of the cluster amplitudes $t_{\mu_n} = t^{ab\cdots}_{ij\cdots}$ and the excitation operators $\tau_{\mu_n} = E_{ai} E_{bj} \cdots$.
As an approximation to the CCSD model, the CC2 model only involves the singly and doubly excited cluster operators.

The left-projection of the Hartree-Fock wave function $\ket{HF}$ and excitation determinant $\ket{\mu_i} = \tau_{\mu_i} \ket{HF}$ onto the CC2 Schr{\"o}dinger equation generates the complete set of equations for the CC2 ground-state energy and amplitudes.
\begin{align} 
    E_{CC2} &= \bra{HF} \hat H + \left[ \hat H, T_2 \right] \ket{HF} \\
    \Omega_{\mu_1} &= \bra{\mu_1} \hat H + \left[ \hat H, T_2 \right] \ket{HF} = 0 \\
    \Omega_{\mu_2} &= \bra{\mu_2} \hat H + \left[ F, T_2 \right] \ket{HF} = 0
\end{align}
Here $E_{CC2}$ is the CC2 ground-state energy and the residuals $\Omega_{\mu_i}, i = 1,2$ are termed CC vector functions.
$F$ is the Fock operator.

Two equivalent approaches are available for computing excitation energies within the CC2 framework.
The first, known as the linear response (LR) theory,\cite{sekino1984linear,koch1990coupled,hald2000linear} obtains them by solving the eigenvalue equation associated with the unsymmetric Jacobian matrix $\boldsymbol{A}$ of the vector function $\Omega_{\mu_i}$
\begin{equation}
    A_{\mu_i \nu_j} = \frac {\partial \Omega_{\mu_i}} {\partial t_{\nu_j}} =
    \begin{pmatrix}
        \bra{\mu_1} \left[ \hat H, \tau_{\nu_1} \right] + \left[ \left[ \hat H, \tau_{\nu_1} \right], T_2 \right] \ket{HF} & \bra{\mu_1} \left[ \hat H, \tau_{\nu_2} \right] \ket{HF} \\
        \bra{\mu_2} \left[ \hat H, \tau_{\nu_1} \right] \ket{HF} & \delta_{\mu_2 \nu_2} \epsilon_{\mu_2}
    \end{pmatrix}
\end{equation}
The doubles-doubles block is diagonal with the orbital energy difference $\epsilon_{\mu_2}$
\begin{equation}
    \epsilon_{\mu_2} = \epsilon_{aibj} = \epsilon_a - \epsilon_i + \epsilon_b - \epsilon_j
\end{equation}
so the eigenvalue equation of the Jacobian can be solved with the effective singles-singles block $A^{eff}_{\mu_1 \nu_1}$
\begin{equation}
    A^{eff}_{\mu_1 \nu_1} = A_{\mu_1 \nu_1} - \frac{A_{\mu_1 \gamma_2} A_{\gamma_2 \nu_1}}{\epsilon_{\gamma_2} - \omega}
\end{equation}
Solutions from the right and the left respectively give the right ($r_{\mu_1}$) and left ($l_{\mu_1}$) eigenvectors with the same excitation energy $\omega$.
\begin{align}
    A^{eff}_{\mu_1 \nu_1} (\omega) r_{\nu_1} &= \omega r_{\mu_1} \\
    l_{\mu_1} A^{eff}_{\mu_1 \nu_1} (\omega) &= \omega l_{\nu_1}
\end{align}
Another approach is the equation-of-motion (EOM) formulation.\cite{geertsen1989equation,comeau1993equation,stanton1993equation,stanton1995perturbative,gwaltney1996simplified,hohenstein2013tensor,haldar2022similarity}
The wave function for the $m$-th excited state is constructed by applying a linear excitation operator $R_m$ to the CC2 ground-state wave function.
\begin{gather}
    \ket{\psi_m} = R_m \ket{\psi_0} \\
    R_m = r^m_0 + R^m_1 + R^m_2 = r^m_0 + \sum^2_{i=1} {r^m_{\mu_i} \tau_{\mu_i}}
\end{gather}
After a sequence of transformations, the final EOM-CC2 equation is derived
\begin{equation}
    \left[ \bar H_N, R_m \right] \ket{HF} = \omega_m R_m \ket{HF}
\end{equation}
The normal-ordered $T_1$-transformed Hamiltonian is defined as
\begin{gather}
    \bar H_N = e^{-T_1} H e^{T_1} - \langle HF \left| e^{-T_1} H e^{T_1} \right| HF \rangle \\
    \left( \mathit{\bar H_N} \right)_{\mu_i \nu_j} =
        \begin{pmatrix}
            0 & \bra{HF} \bar H_N \ket{\nu_1} & \bra{HF} \bar H_N \ket{\nu_2} \\
            0 & \bra{\mu_1} \bar H_N \ket{\nu_1} & \bra{\mu_1} \bar H_N \ket{\nu_2} \\
            0 & \bra{\mu_2} \bar H_N \ket{\nu_1} & \bra{\mu_2} F_N \ket{\nu_2}
    \end{pmatrix}
\end{gather}
The two-by-two block in the lower right corner is equivalent to the Jacobian in the LR theory, and thus gives the same excitation energy.

Due to the non-Hermitian properties of $\bar H_N$, solving the eigenvalue problem from the left side also yields a set of intermediates
\begin{equation}
    L_m = L^m_1 + L^m_2 = \sum^2_{i=1} {l^m_{\mu_i} \tau_{\mu_i}}
\end{equation}
The right and left eigenvectors are biorthonormal and the excitation energy is given as
\begin{gather}
    \bra{HF} L_m R_n \ket{HF} = \delta_{mn} \label{eq:lr_hf} \\
    \omega_m = \bra{HF} L_m \left[ \bar H_N, R_m \right] \ket{HF}
\end{gather}

\subsection*{B. CC2 excited-state analytical gradient}

\hspace{1em} 
The analytical gradient formulation for CC2 excited states, using the Lagrange approach,\cite{helgaker1988analytical, christiansen1998response} parallels that of the ground state.\cite{kohn2003analytic,levchenko2005analytic,feng2019implementation,zhao2025stochastic}
\begin{eqnarray}
    \begin{aligned} 
        L &= E_{CC2} + \sum_{\mu \nu} {l_{\mu} A_{\mu \nu} r_{\nu}} + \bar\omega \left( 1-\sum_{\mu} {l_{\mu} r_{\mu}} \right) \\
        & + \sum_{\mu} {\bar{t}_{\mu} \Omega_{\mu}} + \sum_{pq} {\zeta _{pq} \left( F_{pq} - \delta_{pq} \varepsilon _p \right)} + \sum_{pq} {\omega_{pq} \left( S_{pq} - \delta_{pq} \right)}
    \end{aligned}
    \label{eqn:grad_L}
\end{eqnarray}
The total energy of the target excited state is given by the sum of the CC2 ground-state energy ($E_{CC2}$) and the excitation energy, represented by the first and second terms, respectively.
The third term ensures the biorthonormality of the excitation amplitudes, while the fourth term constrains the ground-state amplitudes.
The final two terms arise from the use of a Hartree-Fock reference and encapsulate the orbital relaxation contributions.

Subsequently, the key point of the problem lies in solving for the four Lagrange multipliers, $\bar\omega$, $\bar t_{\mu}$, $\zeta _{pq}$ and $\omega_{pq}$.
Varying the Lagrangian with respect to $l_{\mu}$ or $r_{\mu}$ yields $\bar\omega = \omega$.
\begin{align} 
    \frac {\partial L} {\partial l_{\mu}} &= \sum_{\nu} {A_{\mu \nu} r_{\nu}} - \bar\omega r_{\mu} = 0 \\
    \frac {\partial L} {\partial r_{\nu}} &= \sum_{\mu} {l_{\mu} A_{\mu \nu}} - \bar\omega l_{\nu} = 0
\end{align}
The detailed solution for $\bar{t}_{\mu}$ was derived by H{\"a}ttig $et$ $al$.\cite{hattig2002transition} through differentiation of the Lagrangian $L$ with respect to $t_{\mu}$
\begin{equation}
    \frac {\partial L} {\partial t_{\nu}} = \frac {\partial E_{CC2}} {\partial t_{\nu}} + \sum_{\mu \gamma} {l_{\mu} \frac {A_{\mu \gamma}} {t_{\nu}} r_{\gamma}} + \sum_{\mu} {\bar{t}_{\mu} \frac {\Omega_{\mu}} {t_{\nu}}} = 0
    \label{eqn:pLpt}
\end{equation}
Defining
\begin{align} 
    \eta_{\nu} &= \frac {\partial E_{CC2}} {\partial t_{\nu}} \\
    B_{\mu \gamma \nu} &= \frac {\partial A_{\mu \gamma}} {\partial t_{\nu}}
\end{align}
and then Eq. (\ref{eqn:pLpt}) can be rewritten as
\begin{equation}
    \sum_{\mu} \bar {t_{\mu} A_{\mu \nu}} = - \left( \eta_{\nu} + \sum_{\mu \gamma} {l_{\mu} B_{\mu \nu \gamma} r_{\gamma}} \right)
\end{equation}

The rest two Lagrange multipliers $\zeta _{pq}$ and $\omega _{pq}$ are derived from the variational condition with respect to the MO coefficients $C$
\begin{equation}
    \frac {\partial L} {\partial C_{\mu q}} = 0
\end{equation}
which is known as Z-vector equation.\cite{gerratt1968force,handy1984evaluation,christiansen1998integral}
The two indices $\mu$ and $p$ are defined in AO basis and MO basis, respectively.
For consistency, the index $\mu$ is transformed back to the MO basis through\cite{kohn2003analytic, levchenko2005analytic}
\begin{equation}
    \sum_{\mu} {C_{\mu p} \frac {\partial L} {\partial C_{\mu q}}} = 0
    \label{eqn:zvector}
\end{equation}

Using the unrelaxed one- $(\bar\gamma_{pq})$ and two-body $(\bar\gamma^{pq}_{rs})$ reduced density matrices (RDMs) (see Appendix), the CC2 Lagrangian can be expressed in a compact form
\begin{equation}
    L = \sum_{pq} {\bar\gamma_{pq} h_{pq}} + \sum_{pqrs} {\bar\gamma^{pq}_{rs} (pq|rs)} + \sum_{pq} {\zeta_{pq} \left( F_{pq} - \delta_{pq} \varepsilon _p \right)} + \sum_{pq} {\omega_{pq} \left( S_{pq} - \delta_{pq} \right)}
    \label{eqn:transL}
\end{equation}
with $h_{pq}$ and $(pq|rs)$ being the one- and two-electron integrals, respectively.

The first derivative of Lagrangian with respect to the nuclear coordinates $x$, the analytical gradient, is constructed as
\begin{equation}
    \frac {\partial L} {\partial x} = \sum_{pq} {\gamma_{pq} h^{[x]}_{pq}} + \sum_{pqrs} {\gamma^{pq}_{rs} (pq|rs)^{[x]}} + \sum_{pq} {\omega_{pq} S_{pq}^{[x]}}
    \label{eqn:pLpx}
\end{equation}
The superscript [$x$] indicates the first derivative with respect to $x$.
The orbital response contributions are incorporated in the final one- ($\gamma_{pq}$) and two-PDMs ($\gamma^{pq}_{rs}$)
\begin{align} 
    \gamma_{pq} &\leftarrow \bar\gamma_{pq} \label{eqn:r1_u} \\
    \gamma_{ia} &\leftarrow \zeta_{ai} \label{eqn:r1_z} \\
    \gamma^{pq}_{rs} &\leftarrow \bar\gamma^{pq}_{rs}\label{eqn:r2_u} \\
    \gamma^{ik}_{ak} &\leftarrow \zeta_{ai} \label{eqn:r2_z}
\end{align}

\subsection*{C. CC2 derivative coupling}

\hspace{1em} 
The Lagrangian for the derivative coupling\cite{kjonstad2024coupled} between states $m$ and $n$ is analogous to that of the gradient.
\begin{eqnarray}
    \begin{aligned} 
        L_{mn} &= \mathcal{O}_{mn} + \sum_{\mu} {\bar\gamma_{\mu} \left( \sum_{\nu} {A_{\mu \nu} r^n_{\nu}} - \omega_n r^n_{\mu} \right)} + \bar\omega \left( 1-\sum_{\mu} {l^n_{\mu} r^n_{\mu}} \right) \\
        & + \sum_{\mu} {\bar{t}_{\mu} \Omega_{\mu}} + \sum_{pq} {\zeta _{pq} \left( F_{pq} - \delta_{pq} \varepsilon _p \right)} + \sum_{pq} {\omega_{pq} \left( S_{pq} - \delta_{pq} \right)}
    \end{aligned}
    \label{eqn:der_L}
\end{eqnarray}
Here, the first term is defined as
\begin{equation}
    \mathcal{O}_{mn} = \bra{\psi_m^L (x_0)} \ket{\psi_n^R (x)}
\end{equation}
The nuclear coordinate of the left state $m$ is fixed at $x_0$ and any derivative acts only on the right state $n$ with a varying $x$.
For notational convenience, the dependence on $x_0$ and $x$ is suppressed in Eq. (\ref{eqn:der_L}).
And the derivative of $L_{mn}$ with respect to $x$ at $x_0$ appears as
\begin{equation}
    F_{mn} \left( x_0 \right) \equiv \frac {d \mathcal{O}_{mn}} {dx} |_{x=x_0} = \bra{\psi_m^L (x_0)} \ket{\frac {d}{dx} \psi_n^R (x) _{|_{x=x_0}}} = \frac {dL_{mn}} {dx} |_{x=x_0} = \frac {\partial L_{mn}} {\partial x} |_{x=x_0}
\end{equation}
The second term specifies the excited-state amplitudes.
The $n$-excited-state energy is given as $\omega_n$
\begin{gather}
    \omega_n = \bra{L_n} \left[ \bar H, R_n \right] \ket{HF}=\sum_{\mu \nu} {l^n_{\mu} A_{\mu \nu} r^n_{\nu}} \label{eqn:omega_n} \\
    \bar H = e^{- T_2 - T_1} H e^{T_1 + T_2}
\end{gather}
The remaining four constraints fulfill an analogous function to those defined in the gradient section.

Still, we need to solve the Lagrange multipliers $\bar\gamma_\mu$, $\bar\omega$, $\bar t_\mu$, $\zeta _{pq}$ and $\omega_{pq}$ first.
The stationarity for $r^n_{\mu}$ is
\begin{equation}
    \frac {\partial L_{mn}} {\partial r_{\sigma}^n\ } = l_{\sigma}^m + \sum_{\mu} {\bar\gamma_\mu A_{\mu \sigma}} - \omega_n \bar\gamma_{\sigma} - \sum_{\mu} {\omega_n \bar\gamma_\mu r_{\mu}^n l_{\sigma}^n} - \bar\omega l_{\sigma}^n = 0
\end{equation}
and this yields $\bar\gamma_\mu$ and $\bar\omega$
\begin{align} 
    \boldsymbol{\bar\gamma}^T &= \frac {\boldsymbol{l}^T_m} {\omega_n - \omega_m} \label{eqn:gamma} \\ 
    \bar\omega &= - \omega_n \boldsymbol{\bar\gamma}^T \boldsymbol{r}_n
\end{align}
The stationarity for $t_{\mu}$ reads
\begin{align} 
    \frac {\partial L_{mn}} {\partial t_{\sigma}} = l_{\sigma}^m r_0^n + \delta_{\sigma,ai} \sum_{bj} {l_{aibj}^m r_{bj}^n} + \sum_{\mu \nu} {\bar\gamma_\mu r_{\nu}^n B_{\mu \nu \sigma}} + \sum_{\mu \sigma} {\bar t_{\mu} A_{\mu \sigma}} = 0 
\end{align}
The initial two terms on the right side of the equation are independent of $\bar\gamma_\mu$ and the solution follows an analogous procedure to the gradient scheme.
The remaining two multipliers are solved through Z-vector equation, as detailed in Eq. (\ref{eqn:zvector}).
The additional contributions arising from $\mathcal{O}_{mn}$ addressed separately.
The explicit expressions for the unrelaxed RDMs can be found in the Appendix.
The specific contribution of $\mathcal{O}_{mn}$ to the Z-vector equation is as follows
\begin{align} 
    \sum_{\mu} {C_{\mu p} \frac {\partial \mathcal{O}_{mn}} {\partial C_{\mu q}}} |_{x=x_0} &= \sum_{\mu} {C_{\mu p} \frac {\partial \bra{L_m} e^{-T} e^T \ket{R_n} |_{x=x_0}} {\partial C_{\mu q}}} \\
    &= \bar\gamma_{pq} \left( l_{\mu}^m, R_{\nu}^n \right)
\end{align}
which equals the 1-RDM from $LAR$ terms.
Its derivative with respect to the nuclear coordinates reads
\begin{gather}
    \frac {\partial \mathcal{O}_{mn}} {\partial x} |_{x=x_0} = \sum_{pq} {\bar\gamma^{mn}_{pq} \bra{\mathit{\psi}_p} \ket{\mathit{\psi}^{[x]}_q}} \\
    S^{R[x]}_{pq} = \bra{\mathit{\psi}_p} \ket{\mathit{\psi}^{[x]}_q}
\end{gather}
The overlap derivatives are not symmetric, and here the right derivative $S^{R[x]}_{pq}$ is available in packages like Psi4,\cite{smith2020psi4} Q-Chem,\cite{epifanovsky2021software} etc.

\subsection*{D. sRI approximation and Laplace transform}

\hspace{1em} 
The RI technique adopts a set of auxiliary basis functions to decouple the four-index ERIs
\begin{align} 
    (\alpha \beta|\gamma \delta) &\approx \sum_{PR} {(\alpha \beta|P) \left[ V^{-1} \right]_{PR} (R|\gamma \delta)} \\
    &= \sum_Q {\left[ \sum_P {(\alpha \beta|P) V^{-1/2}_{PQ}} \right] \left[ \sum_R {V^{-1/2}_{QR} (R|\gamma \delta)} \right]} \\
    &= \sum_Q {B^Q_{\alpha \beta} B^Q_{\gamma \delta}}
    \label{eqn:ri4eri}
\end{align}
The three-index RI tensor $B^Q_{\alpha \beta}$ is given as
\begin{equation}
    B^Q_{\alpha \beta} = \sum_P {(\alpha \beta|P) V^{-1/2}_{PQ}}
\end{equation}

The stochastic RI, abbreviated as sRI, introduces an additional set of stochastic orbitals to further decouple the four-index ERIs.\cite{takeshita2017stochastic}
Each element of these stochastic orbitals \{$\theta^{\xi}$\}, $\xi=1,2,\cdots,N_s$, is randomly generated to be 1 or -1 and they exhibit the following property.
\begin{equation}
    \left\langle \theta \otimes \theta \right\rangle_\xi =\frac{1}{N_s} \sum_{\xi = 1}^{N_s} {\theta ^\xi \otimes \left( \theta ^\xi \right)^T} \approx I
\end{equation}
The number of these stochastic orbitals $N_s$, different from $N_{aux}$, is independent of the system size, and this will be discussed further in a subsequent section.
By inserting the $\left\langle \theta \otimes \theta \right\rangle_\xi$ in the right foremost of Eq. (\ref{eqn:ri4eri}), the sRI realization of four-index ERI is expressed as
\begin{eqnarray}
    \begin{aligned}
        (\alpha \beta | \gamma \delta) &\approx \sum_{QS} {\sum_{PR} {(\alpha \beta|P) V^{-1/2}_{PQ} \left(\left\langle \theta \otimes \theta^T \right\rangle_\xi \right)_{QS} V^{-1/2}_{SR} (R|\gamma \delta)}} \\
        &= \left\langle \sum_P {\left[(\alpha \beta|P) \sum_Q {\left( V^{-1/2}_{PQ} \theta_Q \right)} \right]} \sum_R {\left[ (R|\gamma \delta) \sum_S {\left( V^{-1/2}_{SR} \theta_S \right)} \right]} \right\rangle_\xi \\
        &= \left\langle R^{\xi}_{\alpha \beta} R^{\xi}_{\gamma \delta} \right\rangle_\xi = \frac{1}{N_s} \sum_{\xi = 1}^{N_s} {R^{\xi}_{\alpha \beta} R^{\xi} _{\gamma \delta}}
    \end{aligned}
\end{eqnarray}
The sRI tensor $R^{\xi}_{\alpha \beta}$ contains two indices in AO basis, $\alpha$ and $\beta$, and the original size-dependent index $Q$ in the RI tensor is replaced by a size-independent one $\xi$. 
This formulation allows $R^{\xi}_{\alpha \beta}$ to be treated as a two-rank tensor.
Also, the computational cost of four-index ERIs is reduced from $O(N_{aux} N^4_{AO})$ to $O(N_s N^4_{AO})$, where the constant prefactor $N_s$ is conventionally neglected.

In practice, the sRI approximation to four-index ERIs is always combined with the Laplace transform for the evaluation of MP2 energy-like terms, for example,
\begin{eqnarray}
    \begin{aligned}
        \frac {(ai|bj)} {\epsilon_i - \epsilon_a + \epsilon_j - \epsilon_b} &= - \int_{0}^{\infty} \left\langle R^{\xi}_{ai} R^{\xi}_{bj} \right\rangle_\xi e^{\left( \epsilon_i - \epsilon_a + \epsilon_j - \epsilon_b \right) t} dt \\
        &\approx - \sum^{N_z}_z {w_z \left[ \left\langle R^{\xi}_{ai} R^{\xi}_{bj} \right\rangle_\xi e^{\left( \epsilon_i - \epsilon_a \right) t_z} e^{\left(\epsilon_j - \epsilon_b \right) t_z} \right]} \\
        &= - \sum^{N_z}_z {w_z \left\langle \left[ R^{\xi}_{ai} e^{\left( \epsilon_i - \epsilon_a \right) t_z} \right] \left[ R^{\xi}_{bj} e^{\left( \epsilon_j - \epsilon_b \right) t_z} \right] \right\rangle_\xi} \\
        &= - \sum^{N_z}_z w_z {\left\langle N^{R,\xi}_{ai} N^{R,\xi}_{bj} \right\rangle_\xi}
    \end{aligned}
\end{eqnarray}
The orbital energy denominator is decoupled into the exponential forms $e^{\left( \epsilon_i - \epsilon_a \right) t_z}$.
Combined with the sRI tenor $R^{\xi}_{ai}$, it is contracted into $N^{R,\xi}_{ai}$.
\begin{equation}
    N^{R,\xi}_{ai} = R^{\xi}_{ai} e^{\left( \epsilon_i - \epsilon_a \right) t_z}
\end{equation}
The superscript $R$ indicates the origin of this intermediate and is not an index in the final scaling.
Due to the variety of four-index electron repulsion integral (ERI) expressions within the CC2 formalism, analogous cases will subsequently be introduced using this notation.
The variables $w_z$ and $t_z$ denotes the weights and grid points of the numerical quadrature, respectively.
The $N_z$ is the number of these quadrature points and a value of $N_z = 7$ is selected for modest accuracy.
$N_z$ is also independent of the system size, resulting only in an increased prefactor for the computational scaling.

\subsection*{E. Details of sRI-CC2 formulations}

\hspace{1em} 
The mathematical formalism for the CC2 excited-state gradient closely parallels that of the derivative coupling.
Analogous to the CC2 energy calculation, their time-determining steps lie in the tensor contractions of double amplitudes with four-index ERIs (both usually appear in modified forms), for example,
\begin{eqnarray}
    \begin{aligned}
        \sum_{abj} {\hat t_{ij}^{ab} (bj|ak)} &= \sum_{PQabj} {\frac {2 \hat B^P_{ai} \hat B^P_{bj} - \hat B^P_{bi} \hat B^P_{aj}} {\epsilon_i - \epsilon_a + \epsilon_j - \epsilon_b} B_{bj}^Q B_{ak}^Q} \\
        &= - \sum_{z}^{N_z} {w_z \sum_{PQabj} {\left( 2 N^{\hat B, P}_{ai} N^{\hat B, P}_{bj} - N^{\hat B, P}_{bi} N^{\hat B, P}_{aj} \right)} B_{bj}^Q B_{ak}^Q} \\
        &= - \sum_{z}^{N_z} {w_z \sum_{PQabj} {\left[ 2 \left( N^{\hat B, P}_{bj} B_{bj}^Q \right) N^{\hat B, P}_{ai} - \left( N^{\hat B, P}_{bi} B_{bj}^Q \right) N^{\hat B, P}_{aj} \right]} B_{ak}^Q}
    \end{aligned}
    \label{eqn:t_B}
\end{eqnarray}
In the preceding equation, the expression is reformulated using RI and Laplace transform, leading to a decomposition into a tensor contraction of the three-index RI matrix $\hat B^P_{ai}$ with its Laplace-transformed version
\begin{equation}
    N^{\hat B, P}_{ai} = \hat B^P_{ai} e^{\left( \epsilon_i - \epsilon_a \right) t_z}
\end{equation}
This process employs a strategy of canceling identical indices to achieve a lower computational cost.
The procedure exhibits an asymptotic scaling of $O(N^5)$, dominated by the exchange term, while the Coulomb term scales as $O(N^4)$.
A detailed cost analysis for each step, including a comparison with sRI version, is provided in Table \ref{tab:term_scaling}.

The application of sRI approximation to all four-index ERIs in Eq. (\ref{eqn:t_B}), which we refer to as complete sRI, reduces the computational scaling of both the Coulomb and exchange terms to $O(N^3)$.
\begin{eqnarray}
    \begin{aligned}
        \sum_{abj} {\hat t_{ij}^{ab} (bj|ak)} &= \sum_{abj} {\frac {2 \langle \hat R^{\xi'}_{ai} \hat R^{\xi'}_{bj} \rangle_{{\xi'}} - \langle \hat R^{\xi'}_{bi} \hat R^{\xi'}_{aj} \rangle_{\xi'}} {\epsilon_i - \epsilon_a + \epsilon_j - \epsilon_b} \langle R_{bj}^{\xi} R_{ak}^{\xi} \rangle_{\xi}} \\
        &= - \langle \sum_{z}^{N_z} {w_z \sum_{abj} {\left( 2 N^{\hat R, \xi'}_{ai} N^{\hat R, \xi'}_{bj} - N^{\hat R, \xi'}_{bi} N^{\hat R, \xi'}_{aj} \right)} R_{bj}^{\xi} R_{ak}^{\xi}} \rangle_{\xi \xi'} \\
        &= - \langle \sum_{z}^{N_z} {w_z \sum_{abj} {\left[ 2 \left( N^{\hat R, \xi'}_{bj} R_{bj}^{\xi} \right) N^{\hat R, \xi'}_{ai} - \left( N^{\hat R, \xi'}_{bi} R_{bj}^{\xi} \right) N^{\hat R, \xi'}_{aj} \right]} R_{ak}^{\xi}} \rangle_{\xi \xi'}
    \end{aligned}
\end{eqnarray}
\begin{equation}
    N^{\hat R,\xi'}_{ai} = \hat R^{\xi'}_{ai} e^{\left( \epsilon_i - \epsilon_a \right) t_z}
\end{equation}
Two sets of stochastic orbitals $\xi\xi'$ are used because the two four-index ERIs are independent.
The use of stochastic samplings introduces a standard deviation (S.D.), which can be mitigated by utilizing more stochastic orbitals.
The detailed cost for each step is also shown in Table \ref{tab:term_scaling}.

Our prior study\cite{zhao2025stochastic} shows that the complete sRI is not well-suited for the evaluation of gradient-like properties.
This inadequacy stems from the considerable statistical uncertainty, particularly from the sRI treatment of the Coulomb integrals, which drastically increases the computational prefactor $N_s$ required for acceptable precision.
To mitigate this, we propose a partial sRI scheme that restricts the sRI approximation exclusively to the exchange term in the numerator of the double amplitudes.
\begin{equation}
    \sum_{abj} {\hat t_{ij}^{ab} (bj|ak)} = - \sum_{z}^{N_z} {w_z \sum_{abj} {\left[ 2 \left( N^{\hat B, P}_{bj} B_{bj}^Q \right) N^{\hat B, P}_{ai} - \langle \left( N^{\hat R, \xi'}_{bi} B_{bj}^Q \right) N^{\hat R, \xi'}_{aj} \rangle_{\xi'} \right] B_{ak}^Q}}
\end{equation}
The Coulomb term without sRI retains the $O(N^4)$ scaling, while the exchange term with sRI scales as $O(N^4)$ with a discernible loss of accuracy.
The prefactor $N_s$ remains manageable.
In contrast to the complete sRI, which attains an $O(N^3)$ scaling, the partial sRI formulation achieves an $O(N^4)$ scaling but significantly reduces stochastic errors.
An analysis of both partial sRI and complete sRI, based on empirical data, is presented in the subsequent section.

\begin{table}[htbp]
    \caption{Details of intermediate steps. The $A$, $C$ and $D$ in the first column are just intermediates in the calculations. The prefactor $N_s$ in the costs and $\xi \xi'$ in the intermediates are omitted for brevity. `0' in the index means that all the indices are cancelled, only returning a constant. The calculations from $C$ to $D$ for Coulomb and exchange terms can be combined, and here we leave both for completeness.} 
    \begin{adjustbox}{center} 
        \begin{tabular}{cccccc} 
            \toprule[0.3pt]
            \specialrule{0em}{0.3pt}{0.3pt} 
            \midrule
            \multicolumn{3}{c}{Coulomb term} & \multicolumn{3}{c}{Exchange term} \\ 
            \midrule
            \multicolumn{6}{c}{RI} \\
            \midrule
            Term & Index & Cost & Term & Index & Cost \\
            \midrule
            $N^{\hat B, P}_{bj} B_{bj}^Q \rightarrow A^{PQ}$ & $Pbj \times Qbj \rightarrow PQ$ & $O(N^4)$ & $N^{\hat B, P}_{bi} B_{bj}^Q \rightarrow A^{PQ}_{ij}$ & $Pbi \times Qbj \rightarrow PQij$ & $O(N^5)$ \\
            $A^{PQ} N^{\hat B, P}_{ai} \rightarrow C^Q_{ai}$ & $PQ \times Pai \rightarrow Qai$ & $O(N^4)$ & $A^{PQ}_{ij} N^{\hat B, P}_{aj} \rightarrow C^Q_{ai}$ & $PQij \times Paj \rightarrow Qai$ & $O(N^5)$ \\
            $C^Q_{ai} B_{ak}^Q \rightarrow D_{ik}$ & $Qai \times Qak \rightarrow ik$ & $O(N^4)$ & $C^Q_{ai} B_{ak}^Q \rightarrow D_{ik}$ & $Qai \times Qak \rightarrow ik$ & $O(N^4)$ \\
            \midrule
            \multicolumn{6}{c}{Complete sRI} \\
            \midrule
            Term & Index & Cost & Term & Index & Cost \\
            \midrule
            $N^{\hat R, \xi'}_{bj} R_{bj}^{\xi} \rightarrow A_0 $ & $bj \times bj \rightarrow 0 $ & $O(N^2)$ & $N^{\hat R, \xi'}_{bi} R_{bj}^{\xi} \rightarrow A_{ij} $ & $bi \times bj \rightarrow ij $ & $O(N^3)$ \\
            $A_0 N^{\hat R, \xi'}_{ai} \rightarrow C_{ai}$ & $0 \times ai \rightarrow ai$ & $O(N^2)$ & $A_{ij} N^{\hat R, \xi'}_{aj} \rightarrow C_{ai}$ & $ij \times aj \rightarrow ai$ & $O(N^3)$ \\
            $C_{ai} R_{ak}^{\xi} \rightarrow D_{ik}$ & $ai \times ak \rightarrow ik$ & $O(N^3)$ & $C_{ai} R_{ak}^{\xi} \rightarrow D_{ik}$ & $ai \times ak \rightarrow ik$ & $O(N^3)$ \\
            \midrule
            \multicolumn{6}{c}{Partial sRI} \\
            \midrule
            Term & Index & Cost & Term & Index & Cost \\
            \midrule
            $N^{\hat B, P}_{bj} B_{bj}^Q \rightarrow A^{PQ}$ & $Pbj \times Qbj \rightarrow PQ$ & $O(N^4)$ & $N^{\hat R, \xi'}_{bi} B_{bj}^Q \rightarrow A^{Q}_{ij}$ & $bi \times Qbj \rightarrow Qij$ & $O(N^4)$ \\
            $A^{PQ} N^{\hat B, P}_{ai} \rightarrow C^Q_{ai}$ & $PQ \times Pai \rightarrow Qai$ & $O(N^4)$ & $A^{Q}_{ij} N^{\hat R, \xi'}_{aj} \rightarrow C^Q_{ai}$ & $Qij \times aj \rightarrow Qai$ & $O(N^4)$ \\
            $C^Q_{ai} B_{ak}^Q \rightarrow D_{ik}$ & $Qai \times Qak \rightarrow ik$ & $O(N^4)$ & $C^Q_{ai} B_{ak}^Q \rightarrow D_{ik}$ & $Qai \times Qak \rightarrow ik$ & $O(N^4)$ \\
            \midrule
            \specialrule{0em}{0.3pt}{0.3pt}
            \bottomrule[0.3pt]
        \end{tabular}
    \end{adjustbox}
    \label{tab:term_scaling}
\end{table}

\section*{\Rmnum{3}. RESULTS AND DISCUSSION}

\hspace{1em} 
In this section, we first assess performance of our sRI-CC2 excited-state gradient and derivative coupling programs in accuracy and cost across a range of molecular systems.
Subsequently, we evaluate the required number of stochastic orbitals $N_s$ for two distinct sRI approaches.
Our implementation is integrated within the Q-Chem package.
All the calculations employ the cc-pVDZ basis set.
Given the stochastic nature of the sRI technique, the reported sRI-CC2 results represent an average over ten independent calculations, each initialized with a unique random seed.
The corresponding standard deviations are depicted as error bars in the accompanying figures.
The sRI-CC2 variant is deemed accurate if the standard deviation is sufficiently small (a parameter controllable via $N_s$ and tuned for the desired precision) and exceeds the systematic error (i.e., the deterministic RI-CC2 result falls within the error bar).
Unless otherwise stated, we utilize $N_s=100$ for the partial sRI-CC2 and $N_s=50000$ for the complete sRI-CC2.
The rationale for these selections is discussed in Subsec. C.

The output of both the gradient and derivative coupling tests is a matrix of dimensions $3 \times N_{natom}$ (where $N_{natom}$ is the number of atoms), complicating direct result comparison.
To quantify the overall numerical error, we denote the element-wise error as $\Delta_i$ and employ two metrics: the maximum error ($\Delta_{max}$) and the mean absolute error ($\bar \Delta_{abs}$)
\begin{align} 
    \Delta_{max} &= \max_i {|\Delta_{i}|} \\
    \bar\Delta_{abs} &= \frac {1} {n} \sum^n_{i=1} {|\Delta_{i}|}
\end{align}
The former gives the largest error and the latter indicates the general magnitude of error.
We use these two indicators to measure the errors from our sRI-CC2.

All the calculations are carried out in the high performance computing (HPC) center of Westlake University, utilizing an AMD EPYC 7502 (2.5 GHz) node with 64 computational cores.

\subsection*{A. CC2 excited-state gradient}

\hspace{1em} 
We initially benchmark the accuracy of our RI-CC2 implementation with the program in turbomole\cite{kohn2003analytic,balasubramani2020turbomole}.
Subsequently, we assess the validity of our new sRI-CC2 programs across a broader set of molecular systems.
As summarized in Table \ref{tab:grad_var}, we provide the statistical errors for partial sRI-CC2 and complete sRI-CC2 with respect to RI-CC2.
For partial sRI-CC2 with $N_s=100$, all the $\Delta_{max}$ for gradient are maintained below 0.01 hatree/bohr, and all the $\bar \Delta_{abs}$ below 0.002 hartree/bohr.
In contrast, the complete sRI-CC2 method with $N_s=50000$ exhibits significantly larger statistical fluctuations and its errors for both the gradient and standard deviation are several times greater than those of the partial sRI-CC2.
Although these errors can be reduced by increasing the number of stochastic orbitals, the associated computational prefactor becomes prohibitively large for systems containing hundreds of electrons.
Consequently, the potential advantage of complete sRI-CC2 with a cubic scaling may be realized in larger systems beyond our test range.
In the following sections, we therefore focus our analysis on the performance of the partial sRI-CC2 approach.
\begin{table}[htbp]
    \caption{CC2 excited-state gradient error analyses of various systems (in hartree/bohr). The statistical error takes RI-CC2 data as a reference.} 
    \begin{adjustbox}{center} 
        \begin{tabular}{ccccccccc} 
            \toprule[0.3pt]
            \specialrule{0em}{0.3pt}{0.3pt} 
            \midrule
            \multirow{3}*{Molecule} & \multicolumn{4}{c}{Partial sRI-CC2} & \multicolumn{4}{c}{Complete sRI-CC2} \\
            \specialrule{0em}{1pt}{1pt}
            \cline{2-9} 
            \specialrule{0em}{1pt}{1pt}
            & \multicolumn{2}{c}{Gradient} & \multicolumn{2}{c}{S.D.} & \multicolumn{2}{c}{Gradient} & \multicolumn{2}{c}{S.D.} \\
            \specialrule{0em}{1pt}{1pt}
            \cline{2-9} 
            \specialrule{0em}{1pt}{1pt}
            & $\Delta_{max}$ & $\bar \Delta_{abs}$ & $\Delta_{max}$ & $\bar \Delta_{abs}$ & $\Delta_{max}$ & $\bar \Delta_{abs}$ & $\Delta_{max}$ & $\bar \Delta_{abs}$ \\
            \midrule
            $\rm H_2$  & 0.0006 & 0.0004 & 0.0012 & 0.0010 & 0.0001 & 0.0001 & 0.0024 & 0.0017 \\
            $\rm H_2O$ & 0.0011 & 0.0004 & 0.0049 & 0.0026 & 0.0132 & 0.0054 & 0.0221 & 0.0147 \\
            HF         & 0.0031 & 0.0014 & 0.0055 & 0.0030 & 0.0051 & 0.0045 & 0.0236 & 0.0168 \\
            LiH        & 0.0007 & 0.0004 & 0.0007 & 0.0005 & 0.0006 & 0.0003 & 0.0018 & 0.0013 \\
            LiF        & 0.0022 & 0.0009 & 0.0025 & 0.0015 & 0.0019 & 0.0011 & 0.0158 & 0.0085 \\
            $\rm NH_3$ & 0.0014 & 0.0006 & 0.0037 & 0.0021 & 0.0128 & 0.0057 & 0.0182 & 0.0121 \\
            Benzene	   & 0.0045 & 0.0010 & 0.0050 & 0.0021 & 0.1921 & 0.0303 & 0.3062 & 0.0899 \\
            Furan	   & 0.0079 & 0.0016 & 0.0351 & 0.0058 & 0.1145 & 0.0193 & 0.1356 & 0.0565 \\
            Pyrrole	   & 0.0030 & 0.0008 & 0.0081 & 0.0025 & 0.0823 & 0.0162 & 0.2340 & 0.0791 \\
            Pyridine   & 0.0076 & 0.0014 & 0.0079 & 0.0032 & 0.0819 & 0.0172 & 0.2500 & 0.0789 \\
            \midrule
            \specialrule{0em}{0.3pt}{0.3pt}
            \bottomrule[0.3pt]
        \end{tabular}
    \end{adjustbox}
    \label{tab:grad_var}
\end{table}

\begin{figure}[htbp]
    \centering{}
    \includegraphics[width=8cm]{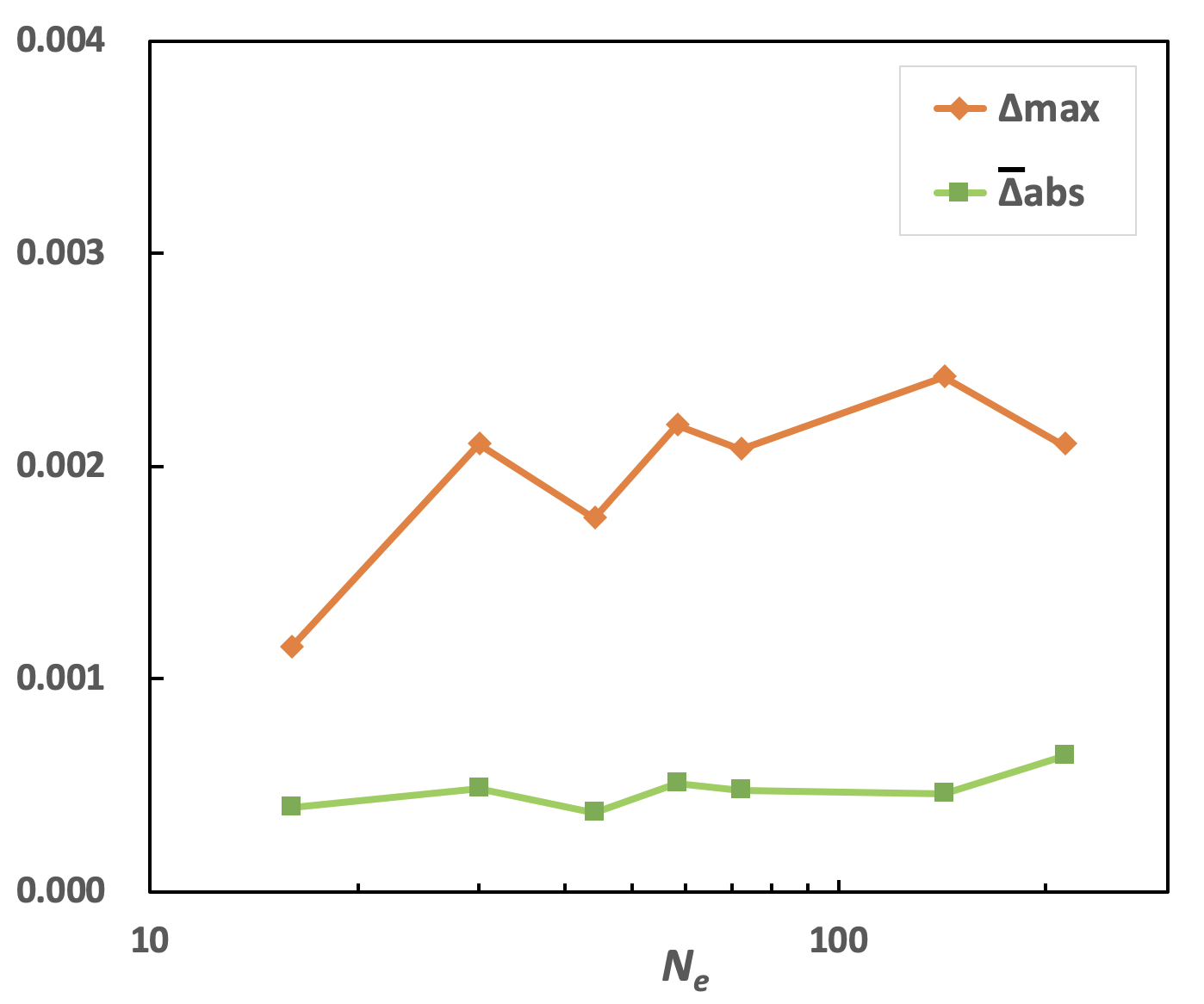} 
    \caption{Error analyses among olefin chains.}
    \label{fig:grad_ole}
\end{figure}

In Table \ref{tab:grad_ole}, we exhibit the partial sRI-CC2 results for a series of (all-$E$)-olefin chains.
Such series extended systems with similar structures and properties provides a convenient framework for evaluating computational costs and error estimation.
A more direct visualization of the data is also shown in Fig. \ref{fig:grad_ole}.
With the increase of the number of electrons $N_e$, the two measures retain the value in a stable range and don't show obvious positive correlation trend.
This indicates that a value of $N_s=100$ is sufficient for all these calculations.
The stochastic error introduced by sRI does not increase with the system size, eliminating the need to adjust $N_s$ for larger molecules.
Therefore, the prefactor $N_s$ proves to be size-independent.
\begin{table}[htbp]
    \caption{Error analyses of partial sRI-CC2 excited gradients among olefin chains (in hartree/bohr).} 
    \begin{adjustbox}{center} 
        \begin{tabular}{ccccc} 
            \toprule[0.3pt]
            \specialrule{0em}{0.3pt}{0.3pt} 
            \midrule
            \multirow{2}*{Molecule} & \multicolumn{2}{c}{Gradient} & \multicolumn{2}{c}{S.D.} \\
            \specialrule{0em}{1pt}{1pt}
            \cline{2-5} 
            \specialrule{0em}{1pt}{1pt}
            & $\Delta_{max}$ & $\bar \Delta_{abs}$ & $\Delta_{max}$ & $\bar \Delta_{abs}$ \\
            \midrule
            $\rm C_2H_4$       & 0.0011 & 0.0004 & 0.0027 & 0.0013 \\
            $\rm C_4H_6$       & 0.0021 & 0.0005 & 0.0040 & 0.0016 \\
            $\rm C_6H_8$       & 0.0018 & 0.0004 & 0.0049 & 0.0017 \\
            $\rm C_8H_{10}$    & 0.0022 & 0.0005 & 0.0043 & 0.0016 \\
            $\rm C_{10}H_{12}$ & 0.0021 & 0.0005 & 0.0038 & 0.0017 \\
            \midrule
            \specialrule{0em}{0.3pt}{0.3pt}
            \bottomrule[0.3pt]
        \end{tabular}
    \end{adjustbox}
    \label{tab:grad_ole}
\end{table}

\begin{figure}[htbp]
    \centering{}
    \includegraphics[width=8cm]{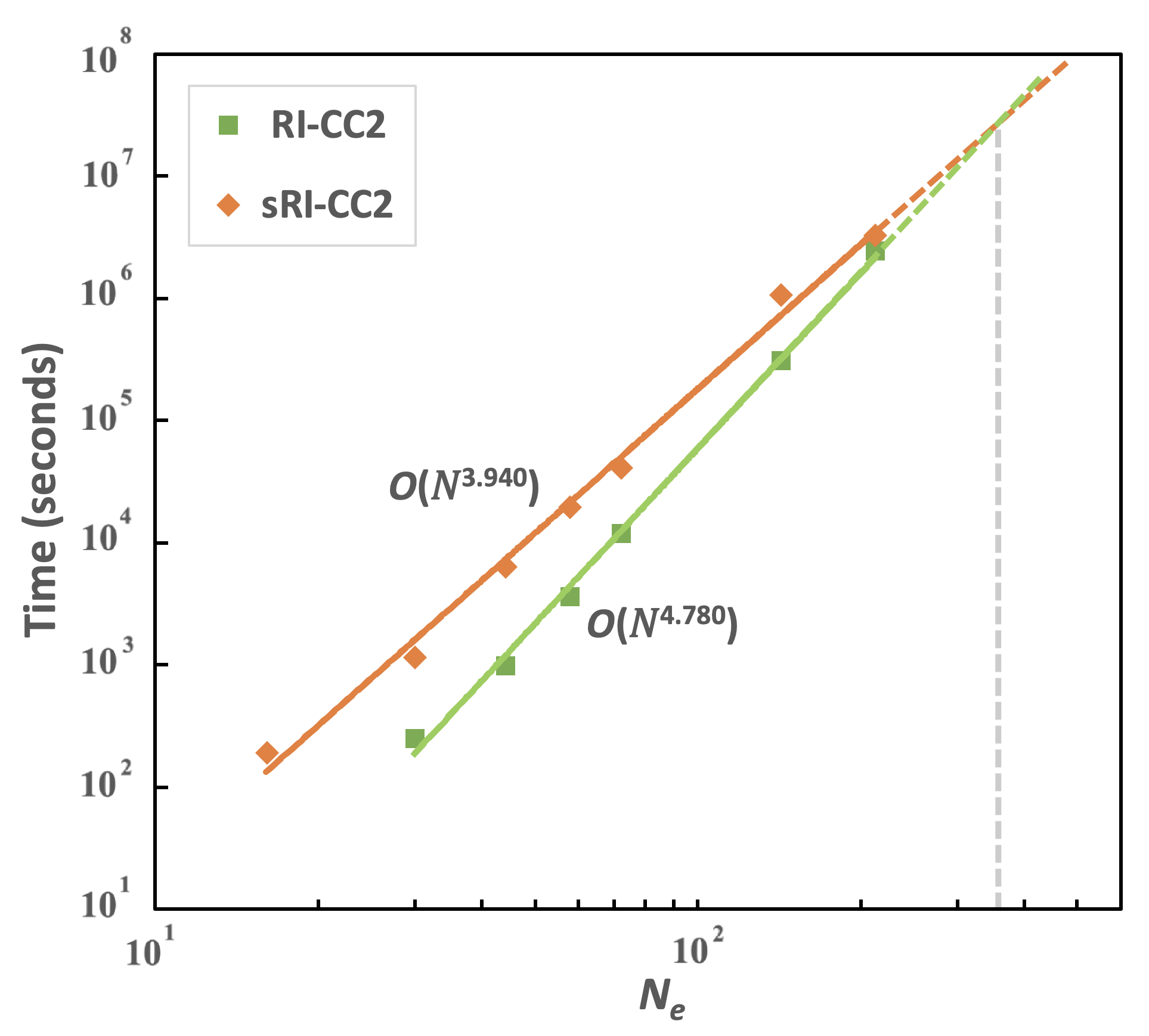} 
    \caption{CPU time for calculations of series olefin chains gradient.}
    \label{fig:grad_time}
\end{figure}

The CPU time for the gradient calculations of series olefin chains is plotted in Fig. \ref{fig:grad_time}.
The RI-CC2 scales as $O(N^{4.78})$, while our partial sRI-CC2 as $O(N^{3.94})$.
The crossover is predicted to occur approximately at $x=400$. Our calculations were terminated near $x=200$ due to the limitation of compute resources.
The RI-CC2 formulation remains advantageous for smaller systems (left of the crossover), while the partial sRI-CC2 method demonstrates a significant performance benefit for larger systems due to its reduced scaling.

\subsection*{B. CC2 derivative coupling}

\hspace{1em} 
The evaluation of CC2 derivative coupling follows a protocol analogous to the one employed for CC2 gradient.
For benchmark references, two established implementations (CCSD and CC2)\cite{tajti2009analytic,kjonstad2023communication} based on the CFOUR and $e^T$ program systems, respectively,\cite{matthews2020coupled,folkestad20201} are in developmental versions and inaccessible.
Consequently, the performance of our sRI-CC2 implementation is preliminarily assessed against CIS results in Q-Chem\cite{fatehi2011analytic} and a CCS implementation by ourselves.

Due to the non-Hermitian of CC2 Jacobian, the absolute value of derivative coupling element encounters a small difference if we exchange the right and left excited states.
Therefore, we average the element in any comparison involving CIS by
\begin{equation}
    \bar F_{mn} = \frac {F_{mn} - F_{nm}} {2}
\end{equation}
The left state $m$ and right one $n$ are configurable in our programs.
In this work, we focus on a pair of nondegenerate states characterized by derivative coupling elements of similar order of magnitude.

Our evaluation begins with a water molecule.
Table \ref{tab:h2o_der_value} presents the largest absolute elements computed at the CIS, CCS, and our RI-CC2 levels.
The values of $\bar F_{41}$ is too small, so we only focus on the performance in $\bar F_{21}$, $\bar F_{31}$ and $\bar F_{51}$.
The two rightmost columns give the relative values with respect to CIS results.
With the CIS results as the references, the error assessment is provided in another manner in Table \ref{tab:h2o_der_err} with the two measures, $\Delta_{max}$ and $\bar \Delta_{abs}$.
The CCS results fit well with the CIS ones.
The RI-CC2 only shows a small difference in the cases of $\bar F_{31}$ and $\bar F_{41}$, while the $\bar F_{21}$ occurs a large error ($\bar\Delta_{abs}=0.1912$).
We attribute this discrepancy to differences between CIS and CC2 in the excited-state amplitude and energy.
To test this hypothesis, we performed a hybrid calculation: the excited-state energy was first converged using CCS, and the derivative coupling was subsequently computed using the CC2 code.
The results are presented in the row, labeled with `CCS$\rightarrow$CC2' in the tables, and show markedly improved agreement with the CIS or CCS values.
These evaluations indicate that the error of our CC2 derivative coupling is manageable in certain cases; however, additional tests are still needed to validate its accuracy.
\begin{table}[htbp]
    \caption{Derivative coupling element of $\rm H_2O$ (in 1/bohr). In the third and fourth columns, we only list the first and second largest values of derivative coupling matrix for comparison. The last two columns show the relative values compared with CIS. `CCS$\rightarrow$CC2' means using the CCS amplitudes in CC2 derivative coupling programs.} 
    \begin{adjustbox}{center} 
        \begin{tabular}{cccccc} 
            \toprule[0.3pt]
            \specialrule{0em}{0.3pt}{0.3pt} 
            \midrule
            \multirow{2}*{Item} & \multirow{2}*{Method} & \multicolumn{2}{c}{Abs largest element} & \multicolumn{2}{c}{Abs relative value} \\
            \specialrule{0em}{1pt}{1pt}
            \cline{3-6} 
            \specialrule{0em}{1pt}{1pt}
            & & 1 & 2 & 1 & 2 \\
            \midrule
            \multirow{4}*{$\bar F_{21}$} & CIS & 2.1581 & 1.1821 & 0.0000 & 0.0000 \\
                                         & CCS & 2.1445 & 1.1753 & 0.0136 & 0.0069 \\
                                         & CC2 & 1.5280 & 0.8700 & 0.6301 & 0.3121 \\
                         & CCS$\rightarrow$CC2 & 2.1693 & 1.1894 & 0.0112 & 0.0073 \\
            \midrule
            \multirow{3}*{$\bar F_{31}$} & CIS & 0.9327 & 0.4546 & 0.0000 & 0.0000 \\
                                         & CCS & 0.9328 & 0.4546 & 0.0001 & 0.0000 \\
                                         & CC2 & 0.9456 & 0.4500 & 0.0129 & 0.0045 \\
            \midrule
            \multirow{3}*{$\bar F_{51}$} & CIS & 0.3391 &        & 0.0000 &        \\
                                         & CCS & 0.3390 &        & 0.0001 &        \\
                                         & CC2 & 0.3382 &        & 0.0009 &        \\
            \midrule
            \specialrule{0em}{0.3pt}{0.3pt}
            \bottomrule[0.3pt]
        \end{tabular}
    \end{adjustbox}
    \label{tab:h2o_der_value}
\end{table}
\begin{table}[htbp]
    \caption{Derivative coupling error analyses of $\rm H_2O$ (in 1/bohr). The statistical error takes CIS data as a reference.} 
    \begin{adjustbox}{center} 
        \begin{tabular}{cccc} 
            \toprule[0.3pt]
            \specialrule{0em}{0.3pt}{0.3pt} 
            \midrule
            Item & Method & $\Delta_{max}$ & $\bar \Delta_{abs}$ \\
            \midrule
            \multirow{3}*{$\bar F_{21}$} & CCS & 0.0136 & 0.0042 \\
                                         & CC2 & 0.6301 & 0.1912 \\
                         & CCS$\rightarrow$CC2 & 0.0253 & 0.0074 \\
            \midrule
            \multirow{2}*{$\bar F_{31}$} & CCS & 0.0001 & 0.0000 \\
                                         & CC2 & 0.0128 & 0.0024 \\
            \midrule
            \multirow{2}*{$\bar F_{51}$} & CCS & 0.0001 & 0.0000 \\
                                         & CC2 & 0.0009 & 0.0002 \\
            \midrule
            \specialrule{0em}{0.3pt}{0.3pt}
            \bottomrule[0.3pt]
        \end{tabular}
    \end{adjustbox}
    \label{tab:h2o_der_err}
\end{table}

Taking the RI-CC2 as the reference, we demonstrate the data from partial sRI-CC2 and complete sRI-CC2 in Table \ref{tab:der_h2o}.
Analogous to CC2 gradient, the latter shows larger errors, while the former reproduces the results of RI-CC2 with adjustable stochastic deviations.
Besides, the $\bar \Delta_{abs}$ for $F_{21}$ is several times greater than the values of $F_{31}$, consistent with the multiple relation of the largest absolute elements in Table \ref{tab:h2o_der_value}.
This observation underscores the importance of benchmarking on systems with coupling elements of comparable magnitude.
More partial sRI-CC2 results are given in Table \ref{tab:der_med}.
\begin{table}[htbp]
    \caption{CC2 derivative coupling error analyses of $\rm H_2O$ (in 1/bohr). The statistical error takes RI-CC2 data as a reference.} 
    \begin{adjustbox}{center} 
        \begin{tabular}{ccccccccc} 
            \toprule[0.3pt]
            \specialrule{0em}{0.3pt}{0.3pt} 
            \midrule
            \multirow{3}*{Item} & \multicolumn{4}{c}{Partial sRI-CC2} & \multicolumn{4}{c}{Complete sRI-CC2} \\
            \specialrule{0em}{1pt}{1pt}
            \cline{2-9} 
            \specialrule{0em}{1pt}{1pt}
            & \multicolumn{2}{c}{Derivative coupling} & \multicolumn{2}{c}{S.D.} & \multicolumn{2}{c}{Derivative coupling} & \multicolumn{2}{c}{S.D.} \\
            \specialrule{0em}{1pt}{1pt}
            \cline{2-9} 
            \specialrule{0em}{1pt}{1pt}
            & $\Delta_{max}$ & $\bar \Delta_{abs}$ & $\Delta_{max}$ & $\bar \Delta_{abs}$ & $\Delta_{max}$ & $\bar \Delta_{abs}$ & $\Delta_{max}$ & $\bar \Delta_{abs}$ \\
            \midrule
            $F_{21}$ & 0.0164 & 0.0076 & 0.0334 & 0.0180 & 0.0767 & 0.0277 & 0.1857 & 0.1368 \\
            $F_{31}$ & 0.0069 & 0.0036 & 0.0280 & 0.0195 & 0.0137 & 0.0078 & 0.0546 & 0.0370 \\
            $F_{51}$ & 0.0026 & 0.0009 & 0.0063 & 0.0043 & 0.0072 & 0.0033 & 0.0189 & 0.0120 \\
            \midrule
            \specialrule{0em}{0.3pt}{0.3pt}
            \bottomrule[0.3pt]
        \end{tabular}
    \end{adjustbox}
    \label{tab:der_h2o}
\end{table}
\begin{table}[htbp]
    \caption{Partial sRI-CC2 derivative coupling error analyses of various systems (in 1/bohr). The error takes RI-CC2 data as a reference.} 
    \begin{adjustbox}{center} 
        \begin{tabular}{cccccc} 
            \toprule[0.3pt]
            \specialrule{0em}{0.3pt}{0.3pt} 
            \midrule
            \multirow{2}*{Molecule} & \multirow{2}*{Item} & \multicolumn{2}{c}{Derivative coupling} & \multicolumn{2}{c}{S.D.} \\
            \specialrule{0em}{1pt}{1pt}
            \cline{3-6} 
            \specialrule{0em}{1pt}{1pt}
            & & $\Delta_{max}$ & $\bar \Delta_{abs}$ & $\Delta_{max}$ & $\bar \Delta_{abs}$ \\
            \midrule
            LiH      & $F_{21}$ & 0.0146 & 0.0087 & 0.0777 & 0.0288 \\
            HCHO     & $F_{21}$ & 0.0054 & 0.0015 & 0.0098 & 0.0049 \\
            Benzene  & $F_{21}$ & 0.0209 & 0.0030 & 0.0283 & 0.0098 \\
            Furan    & $F_{32}$ & 0.0522 & 0.0174 & 0.1717 & 0.0667 \\
            Pyrrole  & $F_{21}$ & 0.0518 & 0.0123 & 0.2995 & 0.0916 \\
            Pyridine & $F_{21}$ & 0.0864 & 0.0256 & 0.2984 & 0.0839 \\
            \midrule
            \specialrule{0em}{0.3pt}{0.3pt}
            \bottomrule[0.3pt]
        \end{tabular}
    \end{adjustbox}
    \label{tab:der_med}
\end{table}

\begin{figure}[htbp]
    \centering{}
    \includegraphics[width=8cm]{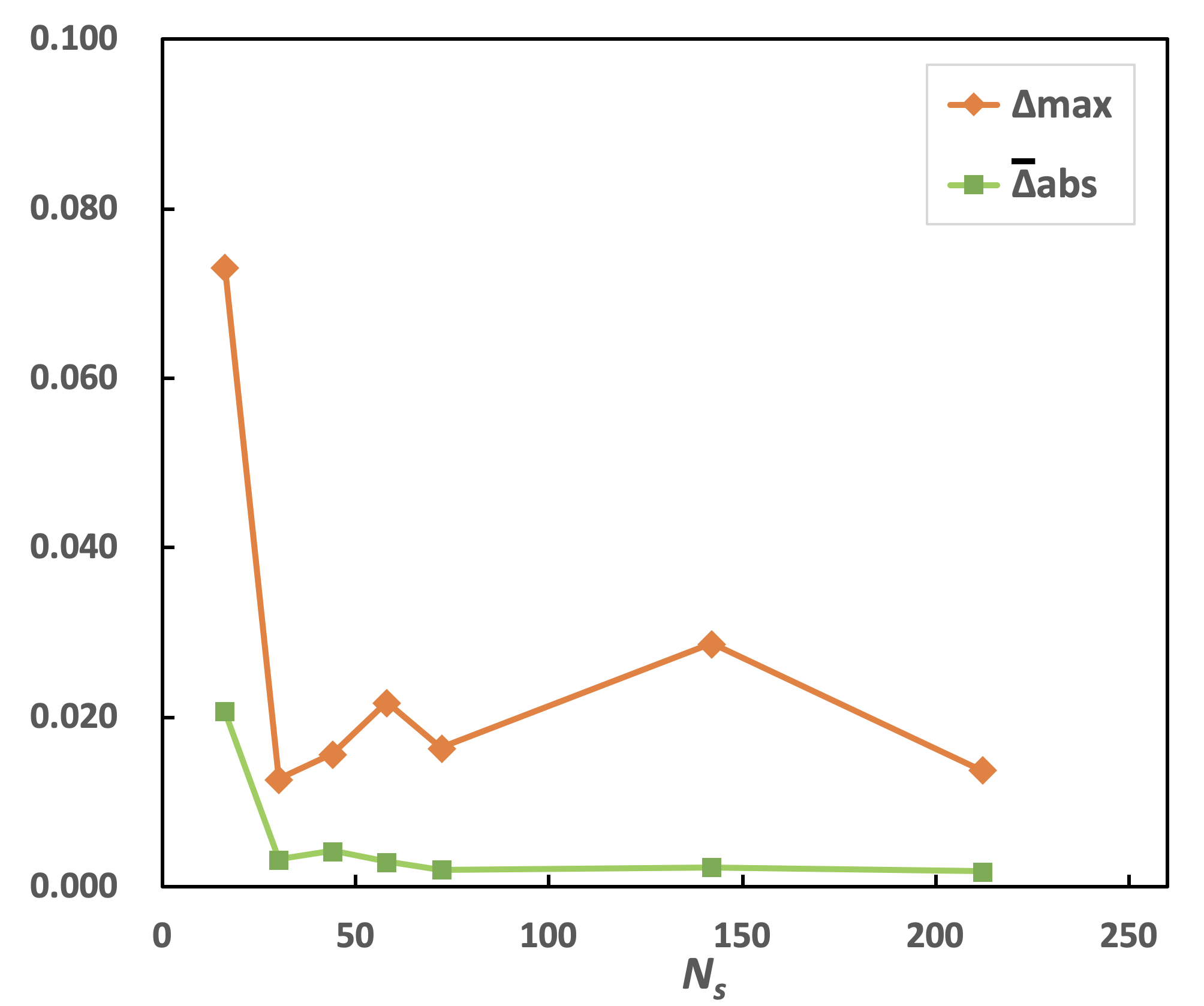} 
    \caption{Error analyses among olefin chains.}
    \label{fig:der_ole}
\end{figure}

The computational cost of partial sRI-CC2 derivative coupling is assessed on the $F_{21}$ among series (all-$E$)-olefin chains in Table \ref{tab:der_ole} and Fig. \ref{fig:der_ole}.
The first row for ethylene shows obvious errors due to its large absolute element and the subsequent series return to normal.
Still, the statistical error is independent of the system size.
Such a result is predictable since the formulation of CC2 derivative coupling is similar to those of gradient.
The error is relatively steady with the increase of system scale and the scaling reduction highlights the performance of sRI in large molecules.
\begin{table}[htbp]
    \caption{Error analyses of partial sRI-CC2 derivative coupling $F_{21}$ among serial olefin chains (in 1/bohr).} 
    \begin{adjustbox}{center} 
        \begin{tabular}{ccccc} 
            \toprule[0.3pt]
            \specialrule{0em}{0.3pt}{0.3pt} 
            \midrule
            \multirow{2}*{Molecule} & \multicolumn{2}{c}{Derivative coupling} & \multicolumn{2}{c}{S.D.} \\
            \specialrule{0em}{1pt}{1pt}
            \cline{2-5} 
            \specialrule{0em}{1pt}{1pt}
            & $\Delta_{max}$ & $\bar \Delta_{abs}$ & $\Delta_{max}$ & $\bar \Delta_{abs}$ \\
            \midrule
            $\rm C_2H_4$       & 0.0732 & 0.0207 & 0.1476 & 0.0746 \\
            $\rm C_4H_6$       & 0.0127 & 0.0032 & 0.0178 & 0.0076 \\
            $\rm C_6H_8$       & 0.0157 & 0.0042 & 0.0206 & 0.0073 \\
            $\rm C_8H_{10}$    & 0.0218 & 0.0030 & 0.0347 & 0.0076 \\
            $\rm C_{10}H_{12}$ & 0.0164 & 0.0020 & 0.0297 & 0.0072 \\
            \midrule
            \specialrule{0em}{0.3pt}{0.3pt}
            \bottomrule[0.3pt]
        \end{tabular}
    \end{adjustbox}
    \label{tab:der_ole}
\end{table}

\begin{figure}[htbp]
    \centering{}
    \includegraphics[width=8cm]{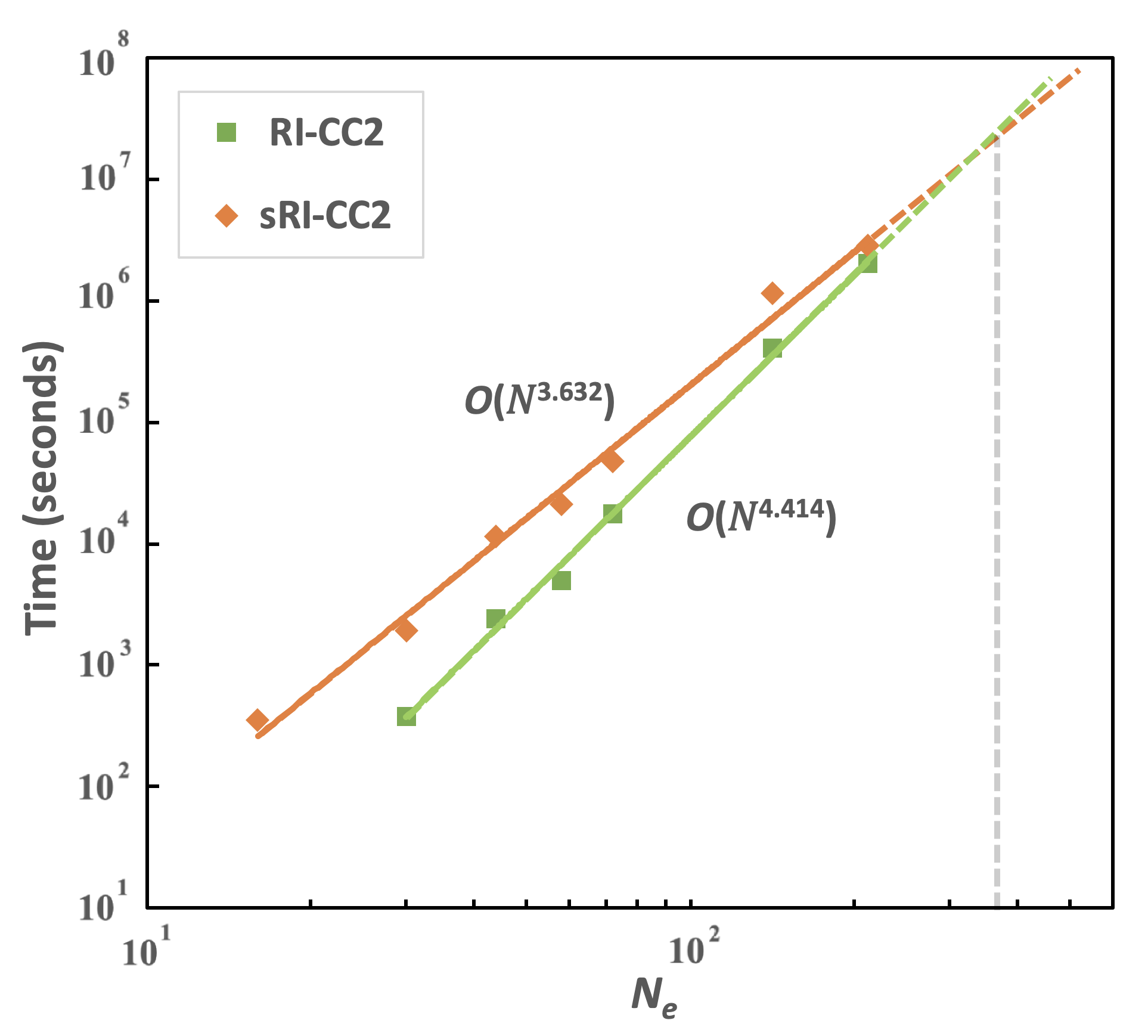} 
    \caption{CPU time for calculations of series olefin chains derivative coupling $F_{21}$.}
    \label{fig:der_time}
\end{figure}

Again, we present the time consumption for various olefin chains in Fig. \ref{fig:der_time}.
The scaling for RI-CC2 is $O(N^{4.41})$, slightly better than $O(N^5)$.
And the computational cost for partial sRI-CC2 is $O(N^{3.63})$, in agreement with its theoretical scaling $O(N^4)$.
The two plots intersect at about $x=400$, beyond which the performance of the partial sRI-CC2 method surpasses that of the RI-CC2 approach.
For larger systems, our partial sRI-CC2 method presents a computationally feasible alternative.

\subsection*{C. Assessment of the number of stochastic orbitals}

\begin{figure}[htbp]
    \centering{}
    \subfigure[]{
        \includegraphics[width=0.45\linewidth]{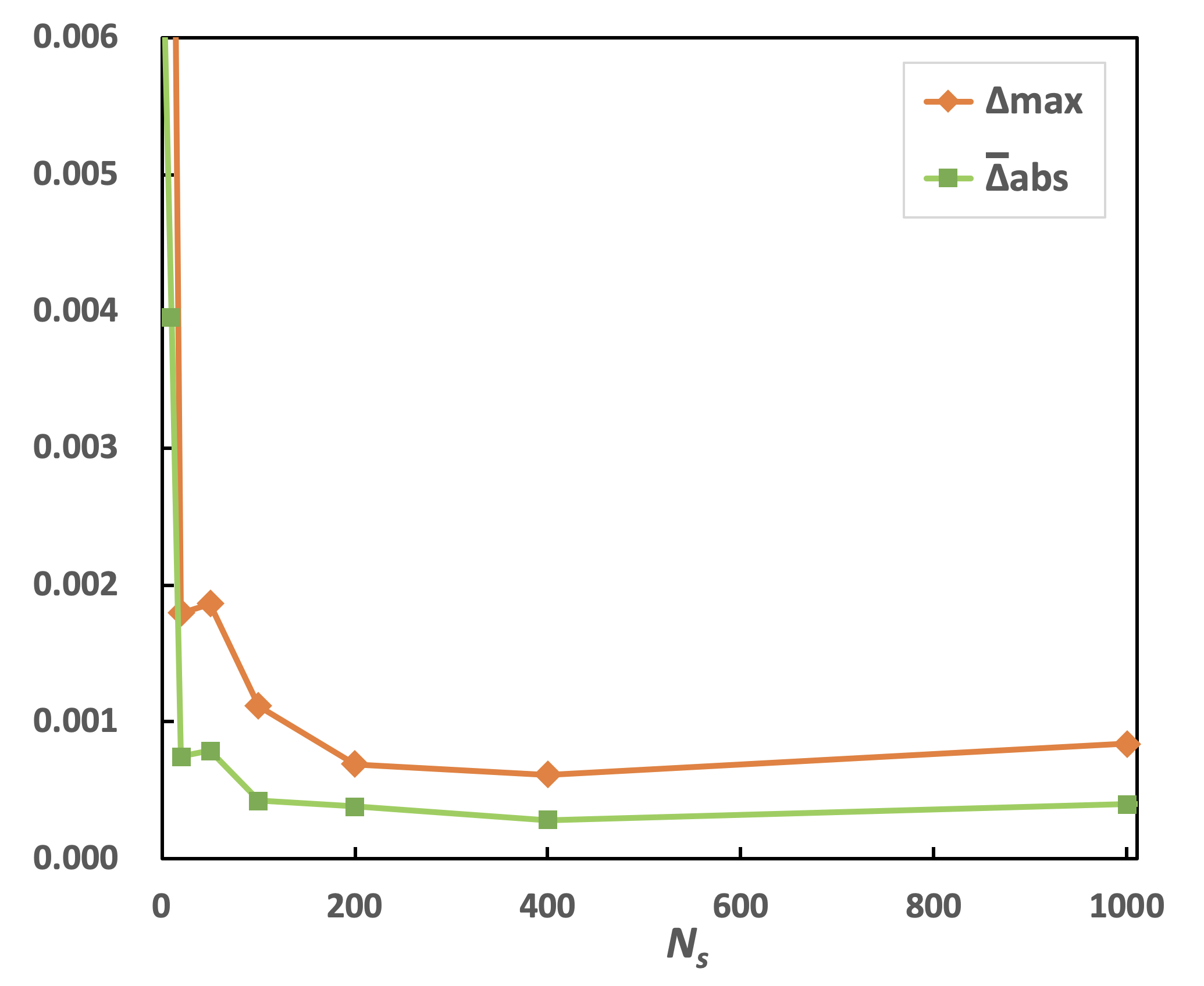}
        \label{fig:grad_psRI_Ns}
    }
    \hfill
    \subfigure[]{
        \includegraphics[width=0.45\linewidth]{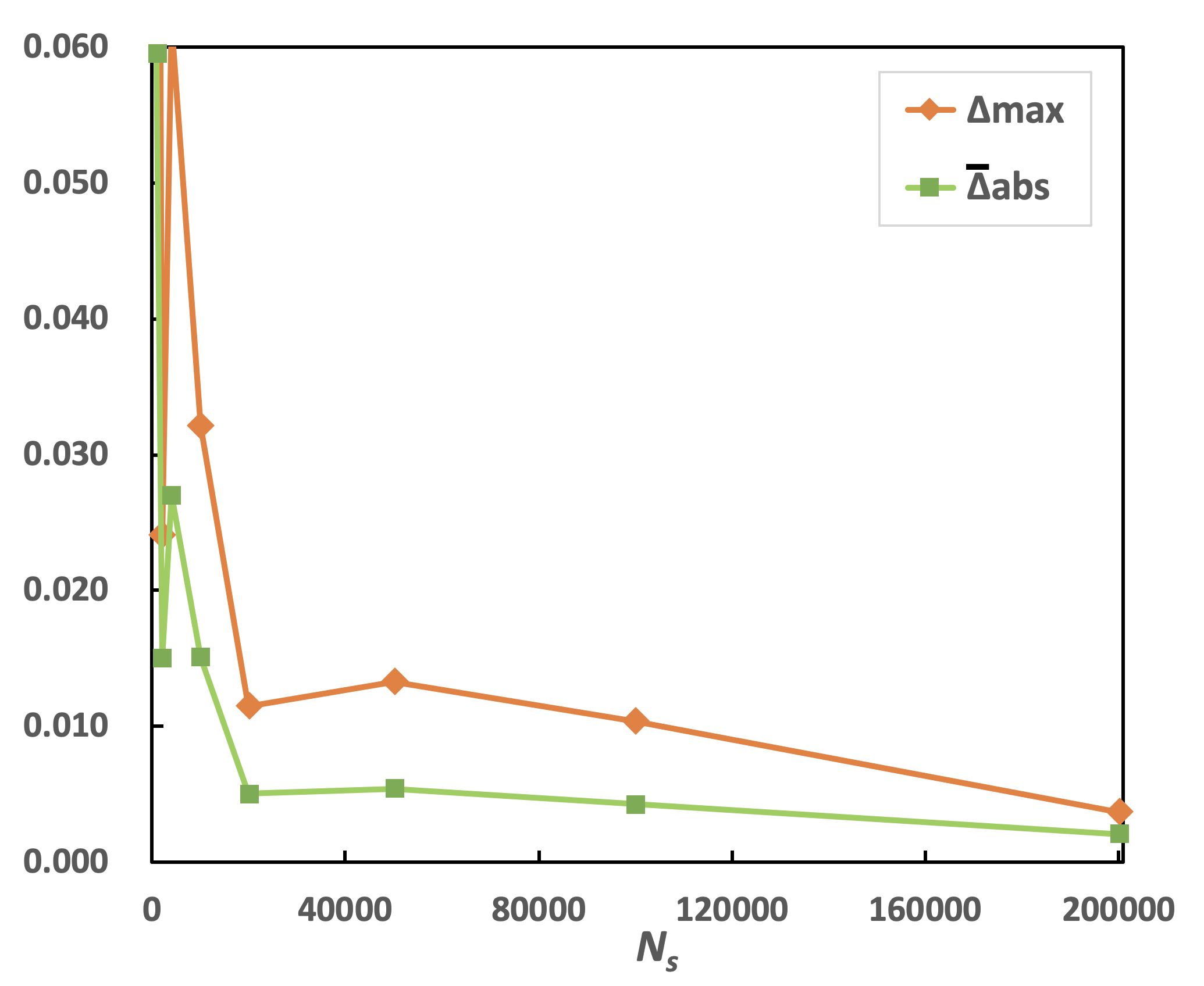} 
        \label{fig:grad_csRI_Ns}
    }
    \caption{Analytical gradient statistical error for $\rm H_2O$ under different number of stochastic orbitals. The left subfigure is from partial sRI-CC2 and the right one from complete sRI-CC2.}
    \label{fig:grad_Ns}
\end{figure}

\begin{figure}[htbp]
    \centering{}
    \subfigure[]{
        \includegraphics[width=0.45\linewidth]{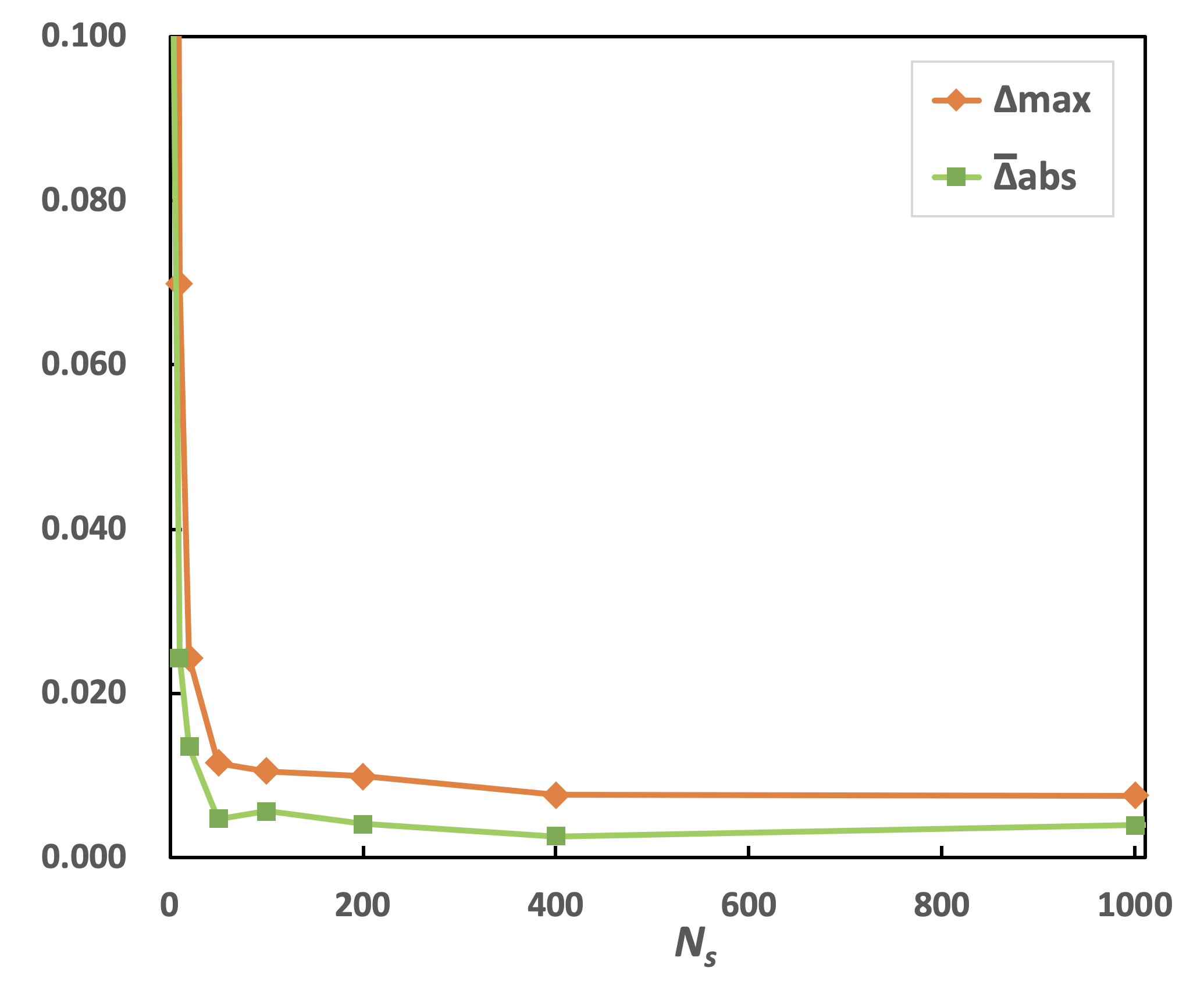}
        \label{fig:der_psRI_Ns}
    }
    \hfill
    \subfigure[]{
        \includegraphics[width=0.45\linewidth]{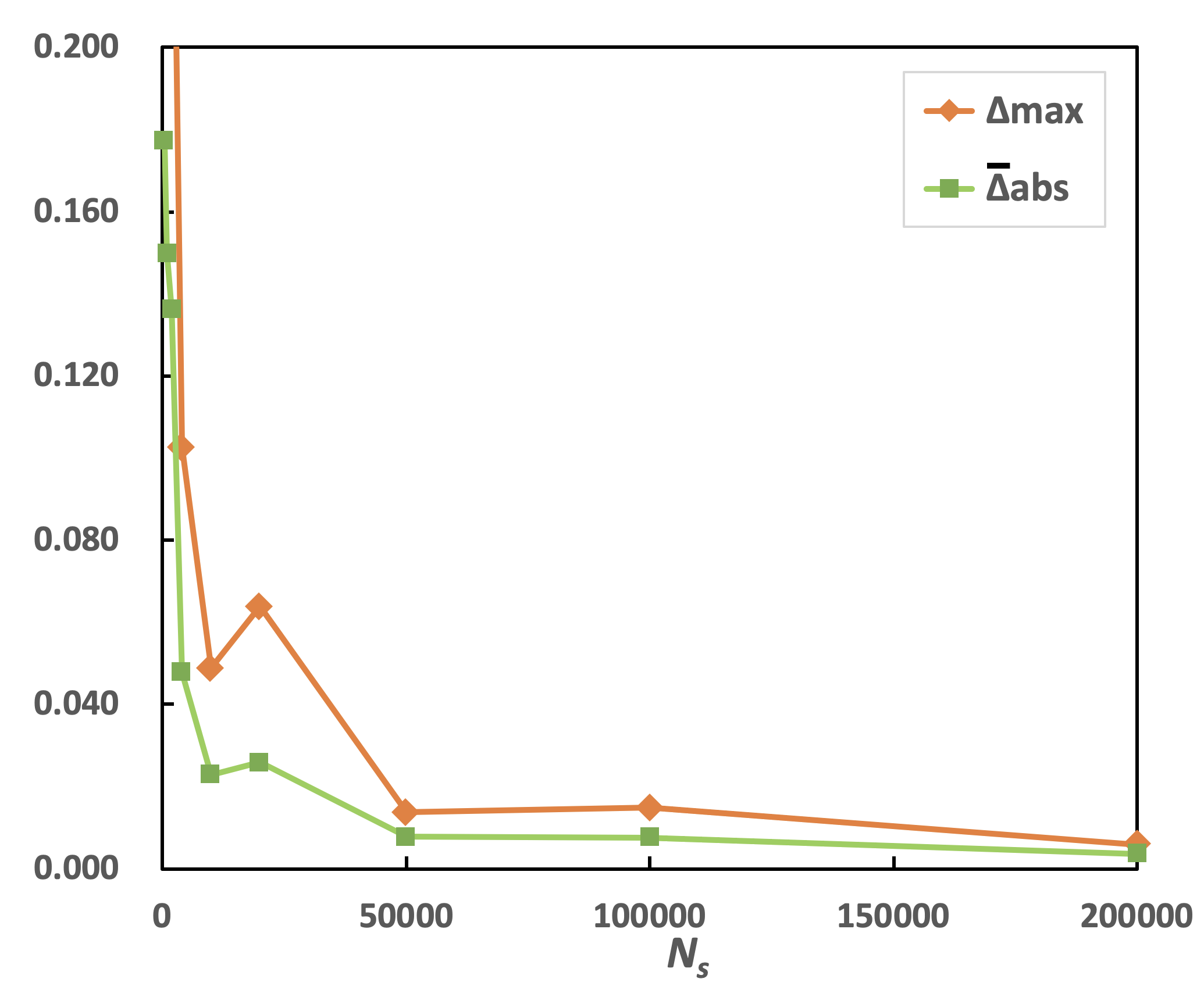} 
        \label{fig:der_csRI_Ns}
    }
    \caption{Derivative coupling statistical error for the $F_{21}$ of $\rm H_2O$ under different number of stochastic orbitals. The left subfigure is from partial sRI-CC2 and the right one from complete sRI-CC2.}
    \label{fig:der_Ns}
\end{figure}

\hspace{1em} 
In this section, we outline the procedures for selecting an optimal number of stochastic orbitals by evaluating the numerical error across a range of $N_s$ values.
In Figs. \ref{fig:grad_Ns} and \ref{fig:der_Ns}, we respectively measure the CC2 gradient and derivative coupling of a water molecule.
With RI-CC2 as the reference, the left subfigure is for partial sRI-CC2 and the right one for complete sRI-CC2.
We notice that all curves decrease sharply with the increase of $N_s$, then tend to be flat, and will eventually converge to zero at large $N_s$ in theory.
A large $N_s$ leads to a smaller stochastic error, but also introduces a large prefactor.
For a compromise between accuracy and cost, we adopt the $N_s$ near the turning point among our tests, $N_s=100$ for partial sRI-CC2 and $N_s=50000$ for complete sRI-CC2.
Beyond these points, the marginal gain in accuracy may be outweighed by the substantial increase in computational cost.

\section*{\Rmnum{4}. CONCLUSIONS}

\hspace{1em} 
In this paper, we apply sRI approximation to CC2 model for the efficient calculations of excited-state analytical gradient and derivative coupling.
While a full sRI implementation reduces the formal computational scaling from $O(N^5)$ to $O(N^3)$, it introduces a large standard noise.
To mitigate this, we introduce a modified approach that applies the sRI approximation selectively to the exchange term.
This hybrid scheme achieves an $O(N^4)$ scaling with markedly improved accuracy.
The sRI-CC2 is particularly advantageous for large molecular systems with hundreds or even thousands of electrons.

The present work constitutes a crucial advancement within our developing sRI-CC2 framework
A primary direction for future research involves extending the sRI formalism to the SCC2 method to calculate conical intersections and realize large-scale dynamic simulations.
The integration of a robust electronic structure method with nonadiabatic dynamics protocols\cite{liu2025scalable} presents a highly promising pathway.

\section*{AUTHOR DECLARATIONS}
\subsection*{Conflict of Interest}
\hspace{1em} 
The authors have no conflicts to disclose.

\section*{ACKNOWLEDGEMENTS}

\hspace{1em} 
We acknowledge the high-performance computing (HPC) service from Westlake University.
W.D. thanks the funding from National Natural Science Foundation of China (No. 22361142829) and Zhejiang Provincial Natural Science Foundation (No. XHD24B0301).
C. L. acknowledges the financial support from the National Natural Science Foundation of China (Nos. 22233001, 22473013) and the Fundamental Research Funds for the Central Universities.
We are grateful for helpful discussions from Fan Wang, Joonho Lee and Qi Ou.

\section*{DATA AVAILABILITY}

\hspace{1em} 
The data that support the findings of this study are available within the article.

\section*{APPENDIX}

\hspace{1em} 
From Eqs. (\ref{eqn:grad_L}) and (\ref{eqn:der_L}), a notable difference of one- and two-RDMs for CC2 excited gradient and derivative coupling, compared to those of ground-state gradient, is the $LAR$ constraint.
The explicit expressions for the additional contributions to these RDMs are provided below.
For the case of CC2 derivative coupling, the $l_{\mu}$ is replace with $\bar \gamma_{\mu}$.
\begin{table}[htbp]
    \caption{Explicit expressions for RDMs.} 
    \begin{adjustbox}{center} 
        \begin{tabular}{ccc} 
            \toprule[0.3pt]
            \specialrule{0em}{0.3pt}{0.3pt} 
            \midrule
            $\bar\gamma_{ji} \leftarrow - \sum_a {l_{ai} r_{aj}}$ & $\bar\gamma_{ab} \leftarrow \sum_i {l_{ai} r_{bi}}$ & \multirow{2}*{$\equiv \bar\gamma^{0}_{pq}$} \\
            $\bar\gamma_{jb} \leftarrow - \sum_{ai} {\left( l_{ai} r_{aj} t_{bi} + l_{ai} r_{bi} t_{aj} \right)}$ & $\bar\gamma_{jb} \leftarrow \sum_{ai} {l_{ai} \hat r^{ab}_{ij}}$ & \\
            \midrule
            $\bar\gamma_{bc} \leftarrow \sum_{aij} {\hat l^{ab}_{ij} r^{ac}_{ij}}$ & $\bar\gamma_{ki} \leftarrow - \sum_{abj} {\hat l^{ab}_{ij} r^{ab}_{kj}}$ & \\
            \midrule
            $\bar\gamma^{pk}_{ql} \leftarrow \bar\gamma^{0}_{pq} I_{kl}$ & $\bar\gamma^{pc}_{qk} \leftarrow \bar\gamma^{0}_{pq} t_{ck}$ & \\
            $\bar\gamma^{bd}_{jl} \leftarrow \sum_{ai} {l_{ai} r_{bj} \hat t^{da}_{li}}$ & $\bar\gamma^{cd}_{kj} \leftarrow - \sum_{ai} {l_{ai} r_{aj} t^{cd}_{ki}}$ & $\bar\gamma^{cb}_{kl} \leftarrow - \sum_{ai} {l_{ai} r_{bi} t^{ca}_{kl}}$ \\
            $\bar\gamma^{ck}_{bl} \leftarrow \sum_{aij} {\hat l^{ab}_{ij} r^{ac}_{ij} I_{kl}}$ & $\bar\gamma^{im}_{kn} \leftarrow - \sum_{abj} {\hat l^{ab}_{ij} r^{ab}_{kj} I_{mn}}$ & \\
            $\bar\gamma^{bs}_{jq} \leftarrow \sum_{ai} {l_{ai} r_{bj} \Lambda^{p}_{aq} \Lambda^{h}_{is}}$ & $\bar\gamma^{bs}_{kj} \leftarrow - \sum_{ai} {l_{ai} r^{ba}_{kj} \Lambda^{h}_{is}}$ & $\bar\gamma^{bc}_{pj} \leftarrow \sum_{ai} {l_{ai} r^{bc}_{ij} \Lambda^{p}_{ap}}$ \\
            $\bar\gamma^{cs}_{pq} \leftarrow \sum_{aibj} {l^{ab}_{ij} r_{ci} \Lambda^{p}_{ap} \Lambda^{p}_{bq} \Lambda^{h}_{js}}$ & $\bar\gamma^{rs}_{kq} \leftarrow - \sum_{ai} {l^{ab}_{ij} r_{ak} \Lambda^{p}_{bq} \Lambda^{h}_{ir} \Lambda^{h}_{js}}$ & \\
            \midrule
            \specialrule{0em}{0.3pt}{0.3pt}
            \bottomrule[0.3pt]
        \end{tabular}
    \end{adjustbox}
    \label{tab:PDM}
\end{table}

\bibliographystyle{IEEEtran} 
\bibliography{main} 

\end{document}